**Article type: Article**

**Unveiling the Role of Dopant Polarity on the Recombination, and Performance of Organic Light-Emitting Diodes**

*Chang-Heon Lee[1], Jeong-Hwan Lee[1,2], Kwon-Hyeon Kim[1] and Jang-Joo Kim*[1]*

[1]Department of Materials Science and Engineering, RIAM, Seoul National University, Seoul 151-744, Republic of Korea
[2]Present address: Department of Material Science and Engineering, Inha University, 100 Inharo, Nam-gu Incheon 22212, Republic of Korea
E-mail: jjkim@snu.ac.kr

**Abstract**

The recombination of charges is an important process in organic photonic devices because the process influences the device characteristics such as the driving voltage, efficiency and lifetime. By combining the dipole trap theory with the drift-diffusion model, we report that the stationary dipole moment ($\mu_0$) of the dopant is a major factor determining the recombination mechanism in the dye-doped organic light emitting diodes when the trap depth ($\Delta E_t$) is larger than 0.3 eV where any de-trapping effect becomes negligible. Dopants with large $\mu_0$ (e.g., homoleptic Ir(III) dyes) induce large charge trapping on them, resulting in high driving voltage and trap-assisted-recombination dominated emission. On the other hand, dyes with small $\mu_0$ (e.g., heteroleptic Ir(III) dyes) show much less trapping on them no matter what $\Delta E_t$ is, leading to lower driving voltage, higher efficiencies and Langevin recombination dominated emission characteristics. This finding will be useful in any organic photonic devices where trapping and recombination sites play key roles.

The recombination of charges refers to a process whereby an electron and a hole are being annihilated and giving off energy. This process produces photons in organic light emitting diodes (OLEDs), but it is a loss mechanism that should be avoided in organic photovoltaics. Therefore, the recombination of charges is an important process in photonic devices, because the process influences the device characteristics such as the driving voltage, efficiency and lifetime. There are two possible recombination processes in dye doped organic semiconductors: Langevin recombination (LR) between a free electron and a free hole, and trap-assisted recombination (TAR) between a trapped charge and an opposite free charge.

Trapping in the dopant is known to affect the recombination mechanism in OLEDs.[1] In addition, this trapping phenomenon is affected by dopant parameters such as trap depth ($\Delta E_t$) and the concentration of dopant.[2,3] Therefore, trap-assisted recombination is known as the dominant mechanism in phosphorescent dye-doped OLEDs (PhOLEDs) because the energy levels of the dopants are located deep compared with the host energy levels with large $\Delta E_t$ and the dopants act as trap sites.[4-12] In contrast, there have been reports that some PhOLEDs with deep trap depths have LR-dominant characteristics[13-18] which cannot be explained based on $\Delta E_t$.[19] It is known that dopant can be considered as a dipole trap and affects charge transport characteristics with energetic disorder arisen from dipole-charge interaction.[20-23] This approach explained the field dependent mobility of molecularly doped system with polar dopant by modifying well known Bässler's Gaussian disorder model. Even with the previously reported results on dipole-charge interaction in literature, its effect on the recombination process has not consider much in organic photonic devices to our best knowledge.

Here, we consider the stationary dipole moment ($\mu_0$) of the dopant as another factor affecting the recombination mechanism and report that $\mu_0$ of the dopant is indeed a major factor influencing the trapping behavior and the recombination mechanism in dye-doped OLEDs. Our experimental results showed that homoleptic Ir-complexes possessing large $\mu_0$ showed trapping-dominant characteristics exhibiting large driving voltage and TAR dominant characteristics, whereas heteroleptic Ir-complexes with small $\mu_0$ lead to very little charge trapping even with $\Delta E_t$, resulting in low driving voltage and LR dominant characteristics. Dopants with larger $\mu_0$ can readily trap charges with stronger Coulomb attraction, which in turn boosts trap-induced characteristics. In addition, drift-diffusion model combined with dipole trap theory was used to investigate how the $\mu_0$ of the dopants affects the recombination mechanism and the device characteristics. The results show that $\mu_0$ of the dopants plays a dominant role rather than $\Delta E_t$ if $\Delta E_t$ is larger than 0.3 eV where any de-trapping effect becomes negligible. Dopants with large $\mu_0$ over 5 Debye (for instance, homoleptic Ir(III) dyes) induce large charge trapping on them. In contrast, dyes with small $\mu_0$ below 2 Debye (e.g., heteroleptic Ir(III) dyes) show much less trapping on them even though the $\Delta E_t$ is much larger than 0.3 eV. The charge trapping and TAR increased the driving voltage significantly to reduce the power efficiency of OLEDs and resultantly the OLEDs doped with heteroleptic Ir(III) complexes resulted in 1.5 times higher power efficiency compared to the homoleptic dye doped OLEDs even though the dyes have almost the same photoluminescence quantum yields (PLQYs).

**Device characteristics of phosphorescent dye doped OLEDs**

Figure 1a shows the schematic diagram of the device structure along with the energy levels of the organic layers. The detailed device structure is ITO(70 nm)/TAPC(75 nm)/TCTA(10 nm)/TCTA:B3PYMPM:Ir dopant(1:1 molar ratio and 8 wt%, 30 nm)/B3PYMPM(45 nm)/LiF(0.7 nm)/Al(100 nm), where TAPC, TCTA and B3PYMPM represent di-[4-(N,N-

ditolyl-amino)-phenyl]cyclohexane, 4,4′,4″-Tris(carbazol-9-yl)triphenylamine and bis-4,6-(3,5-di-3-pyridylphenyl)-2-methylpyrimidine, respectively. The device has an exciplex-forming mixed host to provide a good charge balance and a low injection barrier.[17,24,25] The LUMO of B3PYMPM and the HOMO of TCTA work as the quasi-LUMO and -HOMO levels of the emitting layer. Six different Ir(III) complexes, three heteroleptic and three homoleptic molecules with different energy levels were selected as dopants to investigate the effect of $\mu_0$ and $\Delta E_t$ of the dopants on the recombination mechanism. The heteroleptic dopants are bis(2-phenylpyridine)iridium(III)-acetylacetonate [Ir(ppy)$_2$(acac)], bis(3-methyl-2-phenyl-pyridine)iridium(III)-acetylacetonate [Ir(mpp)$_2$(acac)], and bis(2-phenylpyridine)iridium(III)(2,2,6,6-tetramethylheptane-3,5-diketonate) [Ir(ppy)$_2$(tmd)]. The homoleptic dopants are tris(2-phenylpyridinato-C2,N)iridium(III) [Ir(ppy)$_3$], tris[2-(p-tolyl)pyridine]iridium(III) [Ir(mppy)$_3$], and tris-[2-(1-cyclohexenyl)pyridine]iridium(III) [Ir(chpy)$_3$]. Their chemical structures and energy levels are shown in Figures 1b and c, respectively. The dopants have higher LUMO levels than B3PYMPM and higher HOMO levels than TCTA. Thus, they are expected to behave as hole traps. The device structures were the same for all the dopants to minimize the effect of parameters other than dopant properties.

Figure 2a-b show the current density-voltage *(J–V)* and luminance-voltage *(L–V)* characteristics of the devices, respectively, for the six different dopants. The three OLEDs doped with the heteroleptic dopants show almost similar *J–V* characteristics. However, the other devices doped with the homoleptic dopants exhibited lower current densities than those with heteroleptic dopants at a specific voltage and their *J–V* characteristics are different from each other. Also, the turn-on voltages of the devices doped with the homoleptic dopants are higher than the heteroleptic dopants. The difference in *J–V–L* characteristics between the homoleptic and heteroleptic dopants may be due to the degree of trapping, because the charge

trapping on the dopant lowers $J$ by reducing carrier mobility in the emitting layer.[6,9,14,26,27] Figure 2c shows external quantum efficiency (EQE) against the luminance characteristics. The maximum EQEs were higher 30.0 %, 32.3 %, 30.2 %, 25.8 %, 22.4 % and 23.2 % for Ir(ppy)$_2$(tmd), Ir(ppy)$_2$(acac), Ir(mpp)$_2$(acac), Ir(ppy)$_3$, Ir(mppy)$_3$, and Ir(chpy)$_3$, respectively. These experimental values very well match with maximum achievable EQEs simulated by the classical dipole model considering photoluminescence quantum yield and horizontal dipole ratio (30.5 %, 32.0 %, 30.3 %, 25.9 %, 22.3 % and 22.6 %).[25,28,29] These results show that the devices are well optimized electrically and optically with excellent hole and electron balance. Figure 2d shows the comparison of the operating voltages among the devices. The difference of the driving voltage ($\Delta V$) from the Ir(ppy)$_2$(acac) doped device at same the luminance is plotted against the luminance of the devices. The operating voltage of the devices shows totally different behaviors depending on the type of the dopants. Homoleptic dyes resulted in higher $\Delta V$ than the heteroleptic dyes and the $\Delta V$ of the devices with the homoleptic dyes at 20,000 cd m$^{-2}$ increase three times higher than the values at 10 cd m$^{-2}$, whereas the values of the device with the heteroleptic dyes remain as almost same independent of the luminance. These results are attributed to the difference of the trap characteristics in the two types of the dopants. For the devices with a homoleptic dopant, higher bias should be applied to achieve a specific luminance due to local electric field formed by trapped charges.

The transient EL of the devices clearly confirmed charge trapping in the homoleptic Ir(III) complex-doped devices (Fig. 3). None of the heteroleptic dopant-based devices exhibited any overshoot in the decay curves under reverse bias after turn-off of the electrical pulse. In contrast, all the homoleptic dopant-based devices exhibited overshoots under reverse bias, although the degree of overshoot differed with different dopants. This overshoot is due to the recombination of residual trapped charges in the dopant, which accelerates the process with increasing reverse bias.[18,30,31] Dopants with higher HOMO levels [Ir(mppy)$_3$ and Ir(chpy)$_3$], corresponding to a deeper $\Delta E_t$, resulted in higher overshoots than the device with the lower HOMO level dopant [Ir(ppy)$_3$]. However, the device with the Ir(mpp)$_2$acac having a similar HOMO level with Ir(mppy)$_3$ and Ir(chpy)$_3$ interestingly doesn't exhibit any overshoot. The *J–V–L* characteristics and transient EL measurements indicate that LR is dominant in the devices with the heteroleptic dopants and that TAR is dominant in those with the homoleptic dopants. Still, however, it is unclear why the heteroleptic and homoleptic Ir(III) complex-doped devices give different charge trapping and recombination mechanisms. One needs to notice that the $\Delta E_t$ of the dyes in the dopants are equal to or larger than 0.3 eV.

**Electrical modeling of dye-doped OLEDs**

The difference of the predominant recombination mechanisms in the devices is interpreted based on the different $\sigma_d$ due to the different $\mu_0$ between the two types of dopants. First of all, a modified drift-diffusion model to simulate the electrical properties of the PhOLEDs is developed to clarify these behaviors.[32,33] This model consists of continuity and Poisson's equations. We additionally take into account a continuity equation for trapped holes to simulate the hole-trapping system.

$$\frac{\partial n}{\partial t} = \frac{1}{q}\frac{\partial}{\partial x}\left(\mu_n k_B T \frac{\partial n}{\partial x} + q\mu_n nE\right) - R_L - R_{pt}, \quad (1)$$

$$\frac{\partial p}{\partial t} = -\frac{1}{q}\frac{\partial}{\partial x}\left(\mu_p k_B T \frac{\partial p}{\partial x} - q\mu_p pE\right) - R_L - U_t, \quad (2)$$

$$\frac{\partial p_t}{\partial t} = U_t - R_{pt}, \quad (3)$$

$$\frac{\partial^2 V}{\partial x^2} = -\frac{q}{\varepsilon}(p + p_t - n) \quad (4)$$

where $n$ and $p$ are the densities of free electrons and holes, respectively, $p_t$ is the density of trapped holes, $V$ is the electrostatic potential, $\mu_n$ and $\mu_p$ are the mobilities of electrons and holes, $q$ is the electron charge, $E$ is the electric field, and $\varepsilon$ is the permittivity of the organic material. Measured time-of-flight (TOF) mobilities, summarized in Table 1 were used in the modeling. We considered the field dependence of mobility using the Poole–Frenkel form ($\mu_p = \mu_{p,0} \exp(\gamma \sqrt{E})$). For the simulation, we used 3.5 for the relative permittivity. $R_L$ is the LR rate of a hole and an electron, $R_{pt}$ is the TAR rate between a trapped hole and a free electron and $U_t$ is the sum of trapping and detrapping rate of the hole. Each term is expressed as follows:

$$R_L = \frac{q}{\varepsilon}(\mu_n + \mu_p)np \tag{5}$$

$$R_{pt} = \frac{q}{\varepsilon}\mu_n n p_t \tag{6}$$

$$U_t = \sigma_d v_p \left[ p(N_t - p_t) - p_t(N_{HOMO} - p)\exp\left(-\frac{\Delta E_t}{k_B T}\right) \right] \tag{7}$$

The trapping term in $U_t$ consists of $\sigma_d$, the velocity of holes $v_p$, $p$ and the density of the trapping site $N_t - p_t$. The detrapping consists of $p_t$, the density of empty site $N_{HOMO} - p$, and a Maxwell distribution term with the activation energy (i.e., trap depth, $\Delta E_t$). Note that the trapping of holes in the devices was only taken into account because of the energy level differences between the host and the dopants. Of course, the analysis can easily be extended to electron trapping or the trapping of both electrons and holes for other devices.

The trapping characteristics can be described as a Coulombic interaction between the free charge and $\mu_0$ of the dopant. The trapping strength depends on the polarity of the dopant.

The stronger $\mu_0$ of the dopant is, the more readily it traps an encountered charge. We used the relationship between the $\sigma_d$ of a trap and the magnitude of the $\mu_0$ derived by Belmont.[20]

$$\sigma_d = 0.7\left(1+\frac{\sqrt{3}}{4}\right)\frac{q\mu_0}{32kT\varepsilon}. \quad (8)$$

This relationship describes the interaction between a charge and a dipole; this study examines the case of the interaction between a hole and a phosphorescent dye molecule. Other parameters not mentioned in the text are summarized in Table S1 in Supporting information. The drift-diffusion model predicts the *J–V* characteristics of the devices well as shown in Fig. 2 as the lines, validating the simulation considering $\sigma_d$.

**Langevin recombination against trap-assisted recombination**

The portion of LR in total recombination, $P_L$, of the PhOLEDs was calculated using the drift-diffusion model. $\mu_0$ of the Ir-dopants were calculated with the density functional theory using the program Gaussian09,[34] and were then used to calculate the $\sigma_d$ using equation (8). Geometry optimization was performed using the B3LYP exchange-correlation functional, the LANL2DZ basis set for the Ir atom, and the 6-311G(d) basis set for all other atoms. Table 2 summarizes the calculated $\mu_0$, $\sigma_d$ along with the capture radii, $r$ ($\sigma_d = \pi r^2$) and $\Delta E_t$. The results revealed that the homoleptic dopants have 3–4 times larger $\mu_0$ and $\sigma_d$ than the heteroleptic dopants, respectively. The symmetry of the N-heterocycles exhibiting an electron-deficient region, based on the iridium atom, determine the $\mu_0$ of the Ir complexes.[35] The facial-type homoleptic Ir complexes had three identical main ligands contributing to $\mu_0$ along the C3 axis. In contrast, the $\mu_0$ of heteroleptic Ir complexes is smaller than that of homoleptic

Ir complexes because two N-heterocycles are placed on opposite sides with respect to the iridium atom, which cancels out to reduce the total $\mu_0$.

The drift-diffusion model was used to calculate the LR and TAR rates using $\sigma_d$ of the molecules. We used the mobilities of the TCTA:B3PYMPM blend for the emissive layer (TCTA:B3PYMPM:Ir dopant). The effect of dopant trapping on transport properties was considered by modeling the trapping-detrapping characteristics. The blend layer had lower mobility than a single layer because of reduced charge-hopping sites.[36,37]

$P_L$ were calculated to parameterize the dominance of one recombination mechanism over another in the devices. The ratio of LR was integrated over the emissive layer to calculate $P_L$.[38]

$$P_L = \int \frac{R_L}{R_L + R_{pt}} dx \tag{9}$$

The $P_L$ values of the devices were calculated using the calculated $\sigma_d$ and the $\Delta E_t$ of hole defined by the difference between the HOMO level of the dopant and TCTA. Figure 4 shows the contour plot of $P_L$ values as functions of the capture radius (or $\mu_0$) or $\sigma_d$ and $\Delta E_t$. The calculated $P_L$ values of the heteroleptic dopants were 0.74, 0.59, and 0.63 for Ir(ppy)$_2$(tmd), Ir(ppy)$_2$(acac), and Ir(mppy)$_2$(acac), respectively. For the homoleptic dopants, the $P_L$ values were 0.21, 0.26 and 0.21 for Ir(ppy)$_3$, Ir(mppy)$_3$, and Ir(chpy)$_3$, respectively. The theoretical calculation predicts that LR will be dominant in the heteroleptic complex doped OLEDs and that trap-assisted recombination will dominate in homoleptic complex-doped OLEDs. These results are consistent with the experimental results, showing that the difference in the recombination mechanism is due to different $\sigma_d$ or $\mu_0$ between the homoleptic and heteroleptic Ir(III) complexes. To generalize the effect of the $\mu_0$ on the recombination mechanism, we calculated $P_L$ based on both capture radius and $\Delta E_t$. Figure 4 presents that the effect of $\Delta E_t$ became saturated when the $\Delta E_t$ was > 0.3 eV. This is because the detrapping effect of a hole captured at the dopant decreased markedly with increasing $\Delta E_t$. Above a $\Delta E_t$ of 0.3 eV, the capture radius of the dopant was found to be the major factor that determined $P_L$. This result indicates that $\mu_0$ of the dopant is a crucial factor in determining the recombination mechanism.

**Discussion**

Through the experiments of transient EL and *J-V* characteristics, and the drift-diffusion model considering $\sigma_d$ of the dopant based on dipole trap theory, we showed that the $\mu_0$ of the dopant is the dominant factor dictating the charge transport, efficiency, and recombination mechanism in dye doped OLEDs if the $\Delta E_t$ is larger than 0.3 eV where the detrapping effect minimal. The larger the $\mu_0$ of the dopant is, TAR becomes more dominant over LA. As the traps become shallower than 0.1 eV, LR becomes dominant over trap assisted recombination regardless of how large $\mu_0$ of the dyes is, due to the detrapping effect.

The TAR process in the OLEDs generally causes much higher accumulated charge carriers on the dopants, where the HOMO or LUMO energy levels of the dopants are within the bandgap of the host layer[4-9,11,12], whereas LR attributes to less trapped charge carriers in the OLEDs under the external bias. The accumulated polarons in the dopant state of the OLEDs potentially hinders the effective mobility of the injected charge carriers and annihilates the generated excitons as well as induces unnecessary local-field in the OLEDs during the operation. Therefore, LR be preferred against TAR for high-performance OLEDs[17,18]. In contrast, the spin mixing and resulting conversion of triplet to singlet excitons in recently studied phosphorescent or thermally assisted delayed fluorescent (TADF) dye sensitized fluorescent OLEDs favor TAR on the sensitizers over LR on host molecules Conversely, the recombination between dissociated electrons and holes in photovoltaics works as a loss channel to avoid since triplet harvesting OLEDs utilizing either phosphorescent or TADF dyes have large energy gap between host materials

**Conclusions**

We showed, through experiments and the drift-diffusion model considering the $\sigma_d$ of the dopant based on dipole trap theory, that the $\mu_0$ of the dopant is a major factor dictating the recombination mechanism in dye doped OLEDs if the $\Delta E_t$ is larger than 0.3 eV where the detrapping effect diminishes. LR becomes dominant over trap assisted recombination as the $\mu_0$ of the dye decreases. This can be readily understood because dopants with larger $\mu_0$ have larger $\sigma_d$ and thus easily capture charges passing by. As the traps become shallower than 0.1 eV, LR becomes dominant over trap assisted recombination regardless of how large the $\mu_0$ of the dyes is, due to the detrapping effect.

Although we used phosphorescent OLEDs doped with homoleptic (large $\mu_0$) and heteroleptic (small $\mu_0$) Ir(III) complexes, these findings can be applied generally to fluorescent and thermally assisted delayed fluorescent (TADF) OLEDs and will be useful for phosphorescent and TADF sensitized fluorescent OLEDs where recombination sites play important roles. Drift-diffusion modeling combined with dipole trap theory can be a useful tool for obtaining insights and designing efficient OLEDs.

**Methods**

**Device Fabrication**     Prior to the deposition, the ITO glass were exposed to UV-ozone flux for 15 minutes followed by cleaning with deionized water and boiling IPA. Devices were fabricated under a pressure of $5\times10^{-7}$ Torr. All layers were evaporated thermally and deposited on pre-cleaned patterned ITO electrodes on glass substrates without breaking the vacuum. The active area of the devices is $2\times2$ mm$^2$. All devices were encapsulated with glass lids using an ultraviolet curing resin.

**Device Characterization**     *J-V-L* characteristics were measured with a voltage-source-measure unit (Keithley 237) and a SpectraScan PR650 (Photo Research). Transient EL data were obtained using a pulse generator (Agilent 8114A) and a spectrometer (SpectraPro-300i) connected to a photomultiplier tube (Acton Research, PD-438). Mobilities were measured with time-of-flight measurement equipment (Optel, TOF-401).

**Acknowledgements**

This work was supported by a National Research Foundation grant (2014R1A2A1A01002030) funded by the Ministry of Science, ICT and Future Planning. The authors appreciate Schrödinger® for providing DFT calculation program and Mr. Hyun Shin for his assistance in fabricating OLEDs.

**Authors' contributions**

C.-H. L and J.-H. L equally contributed to this work. C.-H. L performed the experiments, conducted the electrical simulation, analyzed the data and wrote the manuscript. J.-H.L initiated and conducted experiments. initiated the research. K.-H. K calculated the static dipole moments of the Ir complexes using DFT. J.-J. K directed the research and wrote the manuscript.

**References**

1       Wetzelaer, G., Kuik, M., Nicolai, H. & Blom, P. Trap-assisted and Langevin-type recombination in organic light-emitting diodes. *Phys Rev B* **83**, doi:10.1103/PhysRevB.83.165204 (2011).

2       Hoesterey, D. C. & Letson, G. M. The trapping of photocarriers in anthracene by anthraquinone, anthrone and naphthacene. *Jornal of Physics and Chemistry of Solids* **24**, 1609-1615 (1963).

3       Wolf, U., Bassler, H., Borsenberger, P. M. & Gruenbaum, W. T. Hole trapping in molecularly doped polymers. *Chem Phys* **222**, 259-267 (1997).

4       Cleave, V., Yahioglu, G., Barny, P. L., Friend, R. H. & Tessler, N. Harvesting Singlet and Triplet Energy in Polymer LEDs. *Adv Mater* **11**, 285 (1999).

5       Cleave, V. *et al.* Transfer Processes in Semiconducting Polymer-Porphyrin Blends. *Adv Mater* **13**, 44 (2001).

6       Lane, P. A. *et al.* Origin of electrophosphorescence from a doped polymer light emitting diode. *Phys Rev B* **63**, doi:10.1103/PhysRevB.63.235206 (2001).

7       Mäkinen, A. J., Hill, I. G. & Kafafi, Z. H. Vacuum level alignment in organic guest-host systems. *J Appl Phys* **92**, 1598, doi:10.1063/1.1487917 (2002).

8       Gong, X., Ostrowski, J. C., Moses, D., Bazan, G. C. & Heeger, A. J. Electrophosphorescence from a Polymer Guest-Host System with an Iridium Complex as Guest: Forster Energy Transfer and Charge Trapping. *Adv Funct Mater* **13**, 439 (2003).

9       Holmes, R. J. *et al.* Efficient, deep-blue organic electrophosphorescence by guest charge trapping. *Appl Phys Lett* **83**, 3818, doi:10.1063/1.1624639 (2003).

10      Yersin, H. Triplet Emitters for OLED Applications. Mechanisms of Exciton Trapping and Control of Emission Properties.   **241**, 1-26, doi:10.1007/b96858 (2004).

11   Kadashchuk, A., Schols, S., Vakhnin, A., Genoe, J. & Heremans, P. Triplet dynamics and charge carrier trapping in triplet-emitter doped conjugated polymers. *Chem Phys* **358**, 147-155, doi:10.1016/j.chemphys.2009.01.008 (2009).

12   Weichsel, C. *et al.* Storage of charge carriers on emitter molecules in organic light-emitting diodes. *Phys Rev B* **86**, doi:10.1103/PhysRevB.86.075204 (2012).

13   Baldo, M. A., Lamansky, S., Burrows, P. E., Thompson, M. E. & Forrest, S. R. Very high-efficiency green organic light-emitting devices based on electrophosphorescence. *Appl Phys Lett* **75**, 4, doi:10.1063/1.124258 (1999).

14   Uchida, M., Adachi, C., Koyama, T. & Taniguchi, Y. Charge carrier trapping effect by luminescent dopant molecules in single-layer organic light emitting diodes. *J Appl Phys* **86**, 1680, doi:10.1063/1.370947 (1999).

15   Gao, Z. Q. *et al.* High Efficiency and Small Roll-Off Electrophosphorescence from a New Iridium Complex with Well-Matched Energy Levels. *Adv Mater* **20**, 774-778, doi:10.1002/adma.200702343 (2008).

16   Jeon, W. S. *et al.* Ideal host and guest system in phosphorescent OLEDs. *Org Electron* **10**, 240-246, doi:10.1016/j.orgel.2008.11.012 (2009).

17   Park, Y.-S. *et al.* Exciplex-Forming Co-host for Organic Light-Emitting Diodes with Ultimate Efficiency. *Adv Funct Mater* **23**, 4914-4920, doi:10.1002/adfm.201300547 (2013).

18   Lee, J.-H., Lee, S., Yoo, S.-J., Kim, K.-H. & Kim, J.-J. Langevin and Trap-Assisted Recombination in Phosphorescent Organic Light Emitting Diodes. *Adv Funct Mater* **24**, 4681-4688, doi:10.1002/adfm.201303453 (2014).

19   Shi, S. *et al.* Effects of side groups on the kinetics of charge carrier recombination in dye molecule-doped multilayer organic light-emitting diodes. *J. Mater. Chem. C* **3**, 46-50, doi:10.1039/c4tc02414a (2015).

20   Belmont, M. R. The capture cross section of a dipole trap. *Thin Solid Films* **28**, 149-156 (1975).

21   Novikov, S. V. & Vannikov, A. V. Field dependence of charge mobility in polymer matrices. *Chem Phys Lett* **182**, 598 (1991).

22   Dunlap, D. H., Parris, P. E. & Kenkre, V. M. Charage-dipole model for the universal field dependence of mobilities in molecularly doped polymers. *Phys Rev Lett* **77**, 542-545 (1996).

23   Lee, C., Park, S.-K., Yang, M., Lee, N.-S. & Kim, N. Correlation between concentration and disorder of doped trap molecules in space charge field formation. *Chem Phys Lett* **422**, 106-110, doi:10.1016/j.cplett.2006.02.061 (2006).

24   Kim, K. H., Moon, C. K., Lee, J. H., Kim, S. Y. & Kim, J. J. Highly efficient organic light-emitting diodes with phosphorescent emitters having high quantum yield and horizontal orientation of transition dipole moments. *Adv Mater* **26**, 3844-3847, doi:10.1002/adma.201305733 (2014).

25   Kim, S.-Y. *et al.* Organic Light-Emitting Diodes with 30% External Quantum Efficiency Based on a Horizontally Oriented Emitter. *Adv Funct Mater* **23**, 3896-3900, doi:10.1002/adfm.201300104 (2013).

26   Chen, J. & Ma, D. Effect of dye concentration on the charge carrier transport in molecularly doped organic light-emitting diodes. *J Appl Phys* **95**, 5778, doi:10.1063/1.1703834 (2004).

27   Zhang, L. *et al.* A Saturated Red-Emitting Phosphorescent Iridium(III) Complex for Application in Organic Light Emitting Diodes. *J Electrochem Soc* **158**, J243, doi:10.1149/1.3596543 (2011).

28   Chance, R. R., Prock, A. & Silbey, R. in *Advances in Chemical Physics*   1-65 (John Wiley & Sons, Inc., 2007).

29  Wasey, J. A. E. & Barnes, W. L. Efficiency of spontaneous emission from planar microcavities. *Journal of Modern Optics* **47**, 725-741, doi:10.1080/09500340008233393 (2000).

30  Savvateev, V., Yakimov, A. & Davidov, D. Transient electroluminescence from poly(phenylenevinylene)-based devices. *Adv Mater* **11**, 519 (1999).

31  Ma, C. W. *et al.* Time-resolved transient electroluminescence measurements of emission from DCM-doped Alq3 layers. *Chem Phys Lett* **397**, 87-90, doi:10.1016/j.cplett.2004.08.096 (2004).

32  Staudigel, J., Stossel, M., Steuber, F. & Simmerer, J. A quantitative numerical model of multilayer vapor-deposited organic light emitting diodes. *J Appl Phys* **86**, 3895-3910 (1999).

33  Lee, C.-C. *et al.* Electrical and optical simulation of organic light-emitting devices with fluorescent dopant in the emitting layer. *J Appl Phys* **101**, 114501, doi:10.1063/1.2738445 (2007).

34  Gaussian 09 (Gaussian, Inc., Wallingford, CT, USA, 2009).

35  Kim, K. H. *et al.* Phosphorescent dye-based supramolecules for high-efficiency organic light-emitting diodes. *Nat Commun* **5**, 4769, doi:10.1038/ncomms5769 (2014).

36  Groves, C., Koster, L. J. A. & Greenham, N. C. The effect of morphology upon mobility: Implications for bulk heterojunction solar cells with nonuniform blend morphology. *J Appl Phys* **105**, 094510, doi:10.1063/1.3117493 (2009).

37  Koster, L. J. A. Charge carrier mobility in disordered organic blends for photovoltaics. *Phys Rev B* **81**, doi:10.1103/PhysRevB.81.205318 (2010).

38      Kuik, M., Koster, L. J. A., Dijkstra, A. G., Wetzelaer, G. A. H. & Blom, P. W. M. Non-radiative recombination losses in polymer light-emitting diodes. *Org Electron* **13**, 969-974, doi:10.1016/j.orgel.2012.02.009 (2012).

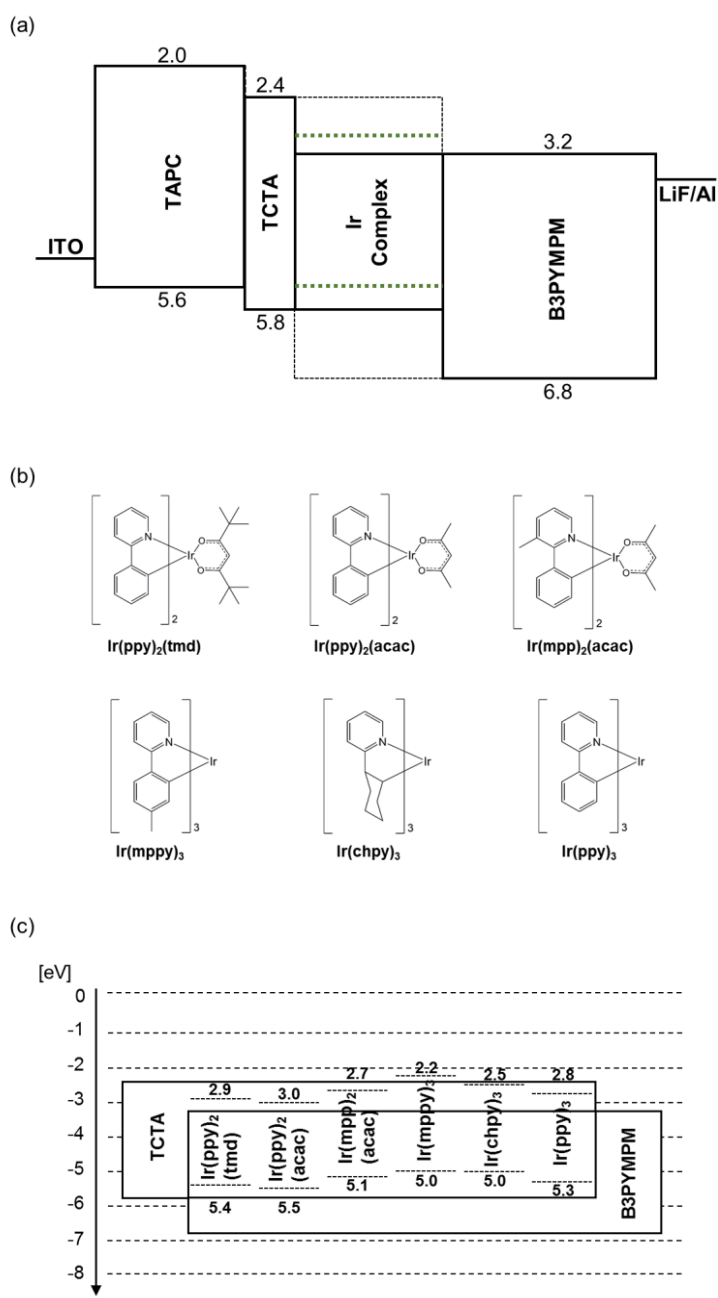

**Figure 1.** (a) Schematic diagram of device structure with HOMO and LUMO levels. (b) Chemical structures and (c) HOMO and LUMO levels of Ir complexes used as dopants in the devices (a). Energies are labeled in eV

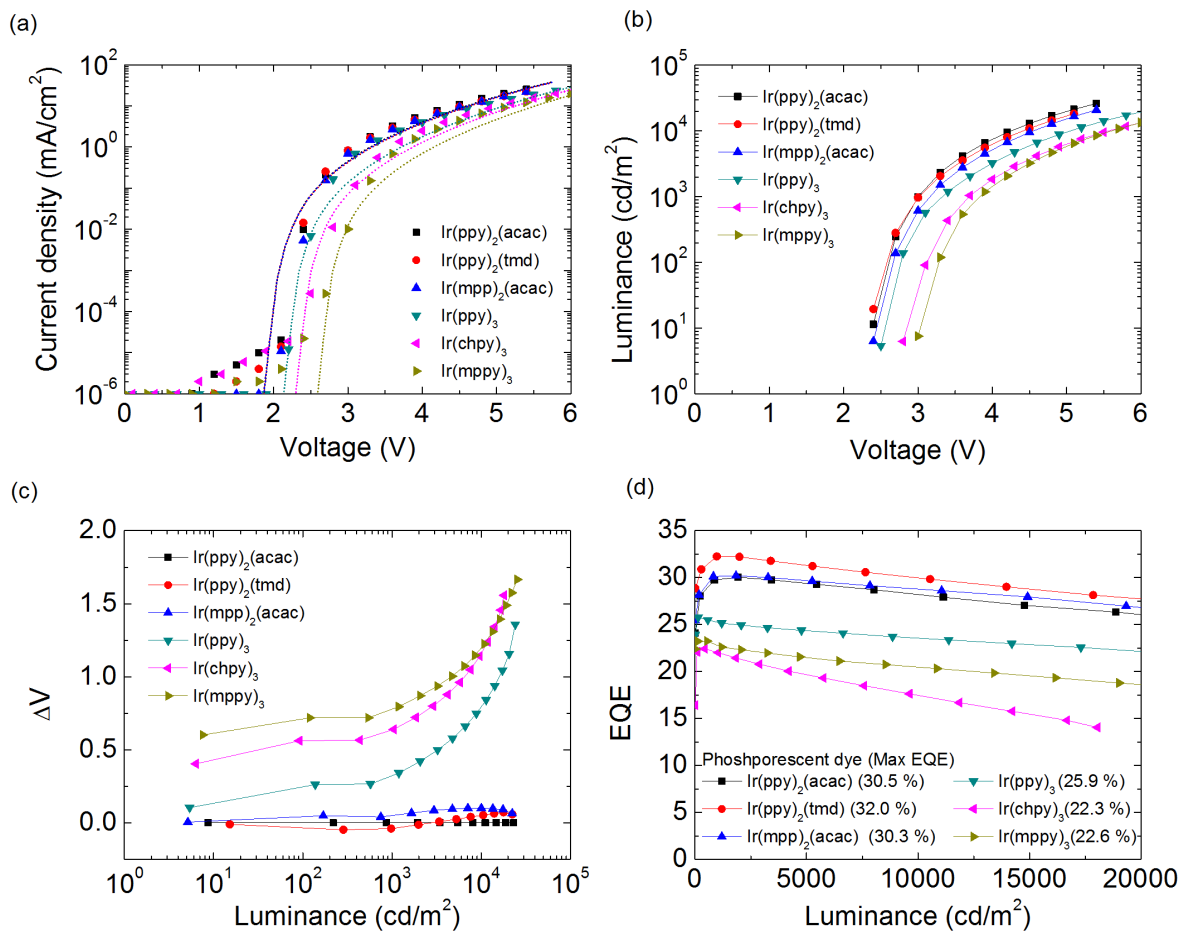

**Figure 2.** (a) Current density-voltage (J-V) , (b) luminance-voltage (L-V), (c) driving voltage difference (ΔV) from the Ir(ppy)2(acac) doped device and (d) external quantum efficiency (EQE) against the luminance characteristics of OLEDs doped with various phosphorescent dyes.

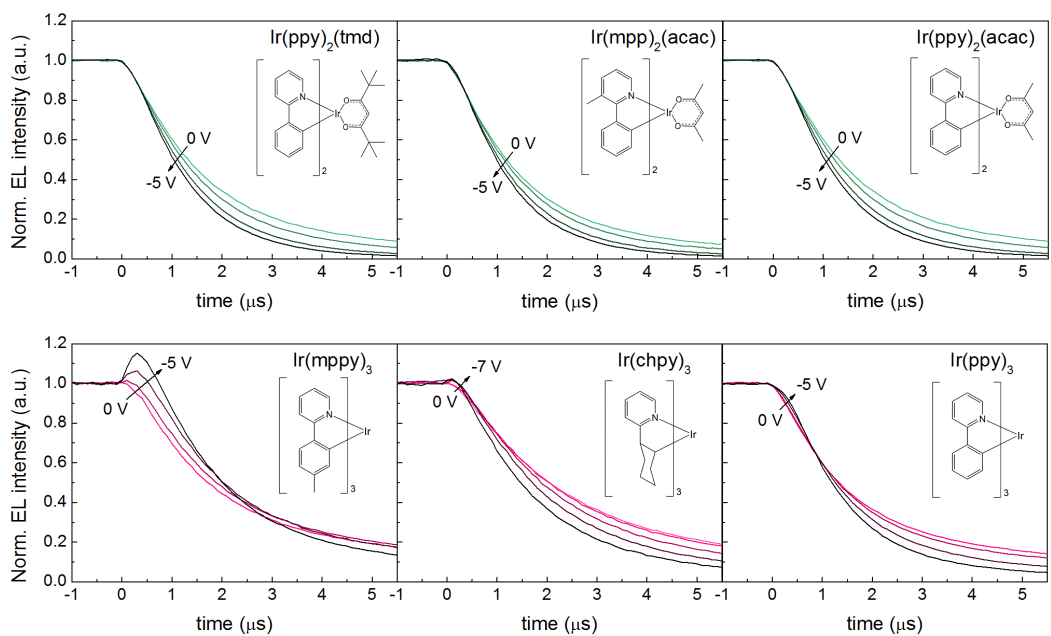

**Figure 3.** Transient electroluminescence characteristics of OLEDs with various Ir dopants. The top three are heteroleptic dopants and the bottom three are homoleptic dopants.

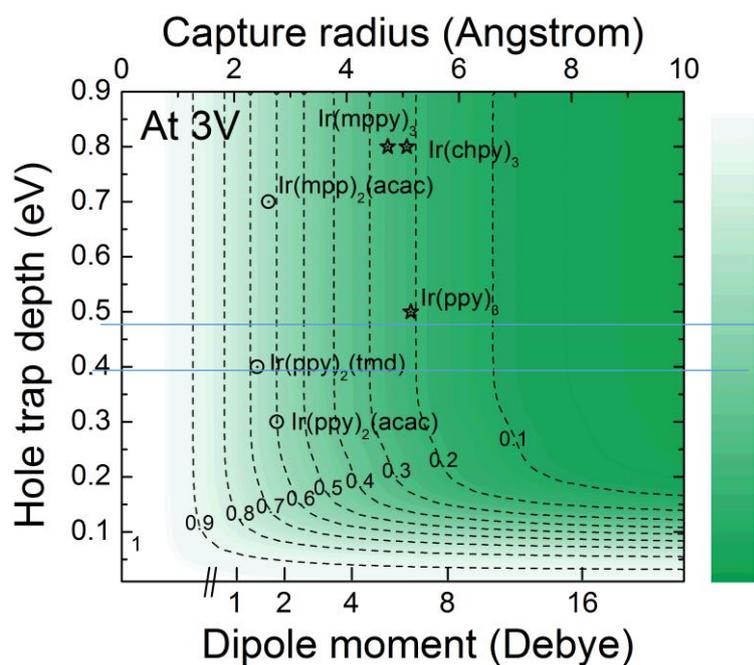

**Figure 4.** Contour plot of simulated portion of LR with functions of $\Delta E_t$ and capture radius of dopants in the device shown in Figure 1a.

**Table 1.** Time-of-flight measured mobilities and Poole-Frenkel constants of the materials used in the model.

| Material | $\mu_p$ (cm$^2$/V) | $\gamma_p$ (cm$^{1/2}$/V$^{1/2}$) | $\mu_n$ (cm$^2$/V) | $\gamma_n$ (cm$^{1/2}$/V$^{1/2}$) |
|---|---|---|---|---|
| TCTA | 5.0×10$^{-5}$ | 2.6×10$^{-3}$ | 1.0×10$^{-8}$ | 2.0×10$^{-3}$ |
| B3PYMPM | 1.0×10$^{-8}$ | 2.0×10$^{-3}$ | 4.5×10$^{-7}$ | 4.5×10$^{-3}$ |
| TCTA:B3PYMPM | 8.6×10$^{-7}$ | 1.8×10$^{-3}$ | 5.0×10$^{-7}$ | 3.0×10$^{-3}$ |

**Table 2.** Capture radii and $\sigma_d$ calculated using equation (8) and $\Delta E_t$ of Ir complexes.

| Heteroleptic dopants | Ir(ppy)$_2$(tmd) | Ir(mpp)$_2$(acac) | Ir(ppy)$_2$(acac) |
|---|---|---|---|
| Dipole moment (Debye) | 1.4 | 1.64 | 1.83 |
| Capture radius (nm) | 0.21 | 0.26 | 0.28 |
| Cross section (nm$^2$) | 0.18 | 0.21 | 0.24 |
| $\Delta E_t$ (eV) | 0.4 | 0.7 | 0.3 |
| Homoleptic dopants | Ir(mppy)$_3$ | Ir(chpy)$_3$ | Ir(ppy)$_3$ |
| Dipole moment (Debye) | 5.38 | 6.18 | 6.36 |
| Capture radius (nm) | 0.47 | 0.51 | 0.51 |
| Cross section (nm$^2$) | 0.70 | 0.81 | 0.83 |
| $\Delta E_t$ (eV) | 0.8 | 0.8 | 0.5 |

**Supplement Materials**

Table S1. Simulation parameters used in the drift-diffusion model

| Parameter | Symbol | Numerical value |
|---|---|---|
| Relative permittivity | $\varepsilon_r$ | 3.5 |
| Temperature | $T$ | 298 K |
| Density of state | $N_0$ | $10^{27}$ m$^{-3}$ |
| Injection barrier | $\phi_B$ | 0.3 eV |